# Competitive Adsorption of Toluene and Water in MFI-type Zeolites

Gavriel Arbiv[a,b,c,‡], Sambhu Radhakrishnan[a,b,‡], Alysson F. Morais[a,b], C. Vinod Chandran[a,b], Dries Vandenabeele[a], Dirk Dom[a,b], Karel Duerinckx[a,b], Christine E. A. Kirschhock[a], and Eric Breynaert[a,b,c,*]

[a] Centre for Surface Chemistry and Catalysis – Characterization and Application Team (COK-KAT), Celestijnenlaan 200F Box 2461, 3001-Heverlee, Belgium.
[b] NMR/X-ray platform for Convergence Research (NMRCoRe), KU Leuven, Celestijnenlaan 200F Box 2461, 3001-Heverlee, Belgium.
[c] Center for Molecular Water Science (CMWS), Notkestraße 85, 22607 Hamburg, Germany.
‡ These authors contributed equally to this work.

KEYWORDS: competitive adsorption, zeolites, NMR spectroscopy, toluene, water

**ABSTRACT:** Competitive adsorption is a major challenge in understanding catalytic activity, selectivity and reaction mechanisms in confined environments such as zeolites. This study investigated competitive adsorption in MFI-type zeolites (ZSM-5) using solid-state NMR, focusing on the interplay between toluene and water. Quantitative $^1$H NMR spectroscopy identified three distinct populations of adsorbed toluene evolving with increasing toluene loading. The adsorption behavior was consistent across a series of samples with Si/Al ratio ranging from 11.5 to 140. Combining 1D and 2D NMR techniques with sample engineering (e.g. pore-blocking) enabled the assignment of the populations to toluene within the zeolite channels, at the pore mouths, and adsorbed on the external crystal surface. Crucially, introducing water to toluene-loaded zeolites caused a partial displacement of toluene from the internal channels, but significant removal from the pore mouths. This displacement occurred even in the highly hydrophobic zeolite (Si/Al = 140), where water still preferentially adsorbed to Brønsted acid sites and silanol species. The results highlight the critical impact that competitive adsorption from solvents, products, or impurities can have on the efficiency and selectivity of zeolite-mediated transformations.

# 1. Introduction

Well-defined microporous frameworks, large surface area, cation exchange capacity, a molecular sized pore system and a very high chemical and hydrothermal stability: these are the properties that make aluminosilicate zeolites indispensable across a wide spectrum of applications, including environmental remediation,[1–3] treatment of water and air,[4] and industrial processes such as refining of petroleum, fuels and hydrocarbons,[5,6] gas purification and separation,[7,8] and fine chemical synthesis.[9–12] Zeolites exhibit varying polarities, depending on topology, framework composition, and on the identity and concentration of extra-framework species and defects. Defect-free purely siliceous zeolites are highly non-polar. Increasing the Al content, the pore wall polarity and density of Brønsted acid sites (BAS) increases.[13–15] The Al content and distribution of T-atoms (e.g., $Si^{4+}$ and $Al^{3+}$) also modulates the zeolite's acidity and its interaction strength with guest species. Al-zoning can be influenced by structure directing agent (SDA)[16], mineralizer,[17,18] etc. in the crystallization and zeolites with identical chemical compositions can still exhibit differences in catalytic and ion-exchange properties. Asides zoning, the Al-content is not necessarily homogeneously distributed within different crystals. This can cause different local catalytical and ion-exchange properties for samples with identical chemical composition, crystal size and morphology.[19,20] Such spatial and compositional heterogeneities lead to complex adsorption profiles and reactivity patterns that are strongly dependent on local properties within the zeolite crystal.[12,19,21,22]

Chemical processes in zeolites occur by interaction of molecules with the framework. Consequently, they are affected by chemical and/or physical adsorption processes. Adsorption in zeolites is governed by steric, enthalpic, and entropic factors, which all are in a dynamic interplay.[23–26] Shape selectivity excludes molecules that cannot enter or diffuse through the pore network due to steric constraints, favoring only those species with dimensions compatible with the pore architecture. Enthalpic selectivity arises from differences (even minor) in adsorption enthalpy between competing guest molecules, and plays a key role, especially at low loadings, where guest–guest interactions are negligible and host-guest interaction dominate adsorption. Host-guest interactions driving adsorption in zeolites include interactions with BAS (chemisorption and physisorption), H-bonding to siloxane bridges and defects, dispersion forces and dipole-dipole interactions.[27–29] At high loadings, the balance among molecular mobility, determining entropy, and favorable molecular interactions, regulates molecular adsorption.[21,30,31] Competitive adsorption can therefore directly affect product selectivity. In hydroalkoxylation of $β$-citronellene with short chain alcohols, increased isomerization and dimerization side-reactions were shown to occur in the absence of alcohol. In those conditions, $β$-citronellene enters the pores of beta-zeolite while in presence of ethanol or water, ethanol and water preferentially occupy the zeolite pores and $β$-citronellene can only adsorb and react at the pore mouths, leading to selective etherification.[32]

Adsorption in zeolites can be studied by X-ray diffraction,[33,34] calorimetry,[35] infrared spectroscopy,[36] quantitative adsorption and solid-state NMR (ssNMR) spectroscopy.[37] A general overview of which information is acquired with each of these techniques is presented in **Table 1**. Among these methods, only X-ray diffraction (XRD) and solid-state NMR spectroscopy provide structural information. Whereas the information from XRD is limited to localization of periodically ordered molecules adsorbed in the zeolite structure, NMR provides quantitative information on the framework protons and adsorbate species and insights on the adsorption behavior and interactions with the framework, even in cases of poor or lacking periodicity.[32] It probes molecular interactions, dynamics, and spatial proximities between adsorbate molecules. Solid-state NMR spectroscopy is therefore one of the most powerful techniques for elucidating the complex interplay of guest molecules within zeolites. *In situ* variable-temperature NMR combined with techniques such as homonuclear and heteronuclear multidimensional correlation NMR (double quantum – single quantum, DQ-SQ), exchange spectroscopy (EXSY), radio frequency driven recoupling (RFDR) and cross polarization heteronuclear correlation (CP-HETCOR) enables to reveal unprecedented details of host–guest interactions, competitive binding, and diffusion phenomena.[32,38–41]

**Table 1**. Comparison of Primary Experimental Techniques for Adsorption Studies in Zeolites

| Technique | Primary Information Obtained | Key Strengths | Key Limitations |
|---|---|---|---|
| **X-ray Diffraction (XRD)** | Crystal structure, phase purity, framework symmetry, localization of periodically ordered guest molecules. | Provides direct structural information on atomic positions; can resolve guest siting. | Averages over the entire crystal; requires high crystallinity; challenging in presence of disorder. |
| **Calorimetry** | Differential/isosteric heat of adsorption ($\Delta H_{ads}$); thermodynamic parameters ($\Delta H^0$, $\Delta S^0$). | Direct, quantitative measure of interaction energy. | Technically demanding; can be affected by slow kinetics or thermal gradients. |
| **Adsorption Isotherms** | Adsorption capacity, selectivity, surface area, pore volume. | Standardized, widely available method for performance evaluation. | Provides macroscopic data; interpretation of mechanism is model-dependent and indirect. |
| **Solid-State NMR** | Local chemical environment (T-sites), nature/quantification of acid sites, framework dynamics, host-guest proximity. | Element-specific; highly sensitive to local structure and dynamics; non-destructive. | Can be insensitive for low-concentration species; complex spectra require advanced techniques for interpretation. |

ZSM5 (Zeolite Socony Mobil 5; framework type MFI illustrated in **Figure 1**) is one of the most intensively studied zeolite materials as it is an important industrial catalyst.[42–46] It finds widespread applications in shape-selective catalysis,[47,48] particularly in aromatic transformation, such as toluene methylation to produce xylene isomers,[19] but also in dealkylation of propylphenol,[23] methanol conversion to hydrocarbons,[49,50] ethane dehydroaromatization,[51] and methane oxidation. This wide range of applications is enabled by its stability, highly tunable range of Si/Al ratio's (from Al-rich Si/Al ratio ~10 to purely siliceous Silicalite-1), and its unique three-dimensional pore architecture with intersecting straight (5.3 × 5.6 Å) and sinusoidal channels (5.1 × 5.5 Å). The intersection of these two channel systems creates cavities of slightly larger dimensions (8.9 Å diameter). This unique dual-pore architecture and the presence of multiple distinct microenvironments is the structural origin that assists to optimize catalytic and shape-selectivity.[19,52,53] The channel dimensions are comparable to the kinetic diameters of many small organic molecules, creating a "tight fit" environment that imposes strong steric constraints on diffusing and reacting species. This confinement is responsible for selectively favoring the formation, adsorption, and transport of specific molecular species by excluding others. In toluene disproportionation for example, para-xylene is kinetically favored due to faster diffusion through narrow pores, despite a higher thermodynamic adsorption coefficient for meta-xylene. Tuning crystallite size and morphology, selective external surface deactivation, and optimized process parameters therefore enable an exceptionally high p-xylene selectivity of more than 90% using MFI zeolites.[54] Recently, a novel mechanism for adsorption of alcohols in high-silica ZSM5 was demonstrated using NMR spectroscopy and later confirmed by calorimetry.[32,55] This new mechanism involves H-bonding to siloxane bridges and explains the room temperature $^{17}$O exchange within zeolite frameworks via equilibration of the zeolite with $^{17}$O enriched water, as well as the methylation of quartz surfaces.[32,56–58]

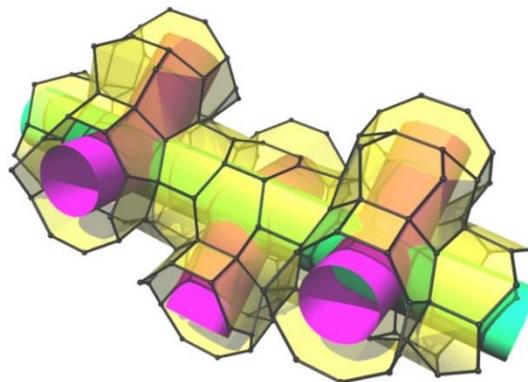

**Figure 1**. Pore structure of an MFI-type zeolite. Straight channels (green) with dimension 5.3 × 5.6 Å and sinusoidal channels (purple) with dimension 5.1 × 5.5 Å intersect in galleries of diameter 8.9 Å.

The pore architecture and wide industrial application of MFI-type zeolites makes detailed studies of their structure–function relationships highly attractive.[12,59] In such studies, it is crucial to recognize that the ZSM-5 framework is not a rigid, static entity. It exhibits a well-documented framework flexibility, most notably a reversible phase transition between a low-temperature monoclinic symmetry (space group P2$_1$/n) and a high-temperature average orthorhombic symmetry (space group Pnma).[11,60] This transition typically occurs in the temperature range of 300 – 350 K and can also be induced by the adsorption of guest molecules, particularly aromatics.[21,31,61,62] The empty Silicalite-1 framework for example exhibits a monoclinic (P2$_1$/n) symmetry. With increasing pore filling the structure adopts an orthorhombic space group (Pnma or P2$_1$2$_1$2$_1$). In the case of benzene adsorption, FT-Raman spectroscopy revealed consecutive phase transitions from P2$_1$/n to P2$_1$2$_1$2$_1$ and finally to Pnma.[63] In a similar experiment using toluene, p-chlorotoluene or chlorobenzene, the lattice rearrangement goes from P2$_1$/n to Pnma and finally to P2$_1$2$_1$2$_1$.[31]

While single component adsorption studies elucidate adsorbent–adsorbate interactions, multicomponent adsorption studies highlight differences in adsorption selectivity of specific pores and can reveal concerted effects, which can severely affect active site occupation and catalytic product selectivity. A combination of confocal laser scanning microscopy and solid-state NMR spectroscopy for example demonstrated that the location where furfuryl alcohol polymerization occurs in ZMS-5, depends on the polarity of the used solvent.[53] Using water as a solvent, furfuryl alcohol oligomerization was directed to the straight channels, while the same reaction occurred in the sinusoidal channels in presence of the less polar 1,4-dioxane.[53] The study clearly demonstrated polarity differences between the two channel types in MFI zeolite, with the preferential location for the solvent and reagent determined by their adsorption preferences.

The commercial significance of toluene transformations in MFI zeolites has motivated extensive research into understanding toluene adsorption, confinement and dynamics within the ZSM-5 zeolite framework.[19,43,44] Crystallographic studies by Nishi et al.[64], and Rodeghero et al.[65] have identified multiple toluene adsorption sites within MFI structures, with preferential occupation of channel intersections at low loadings and subsequent filling of individual channel systems at higher concentrations. Despite its significant implications for catalytic selectivity,[19,52,66] and multiple crystallographic investigations, precise location and distribution of toluene molecules within the MFI channel system remain unclear with conflicting reports describing molecular adsorption sites at industrially relevant concentrations (**Table 2**).[67,68] Results from single-crystal XRD of MFI zeolites loaded with 6.4 toluene molecules per unit cell (uc) identified three distinct toluene sites, one in the sinusoidal channels, and two, mutually excluding, sites in the intersection. The latter hints at dynamic or static disorder.[64] *In situ* synchrotron powder XRD for the same toluene loading located toluene in the straight channels in addition to the intersection. [65] At loadings below the number of channel intersections (4 per unit cell), multiple organic molecules (e.g. phenol, naphthalene, dichlorobenzene) preferentially occupy the cage-like intersections between straight and sinusoidal channels of MFI.[34,69–71] This can be expected to lead to significant displacement effects, altering guest distribution, diffusion, and catalytic outcomes.

**Table 2.** Overview of literature reporting on the positioning of aromatic and aliphatic hydrocarbons adsorbed in MFI zeolites.

| Technique | Adsorbate | Sample condition | Observation | Reference |
|---|---|---|---|---|
| **SCXRD**[A] | Toluene (pure) | Sample: H-MFI Si/Al = 183. Loading: 6.4 molecules per uc | 3 types: TOL1 and TOL2 in the intersection, TOL3 in the sinusoidal channel | Nishi et al.[64] |
| **PXRD**[B] | n-hexane /water (up to 0.16 mg/L) | Sample: H-MFI Si/Al = 140 Loading: 8 molecules per uc | 2 types: HEX1 with four molecules in the intersection, HEX2 with four molecules in the sinusoidal channels. | Rodeghero et al.[72] |
| **FT-Raman** | Toluene (pure) | Sample: siliceous ZSM-5 Loading: 1-4 molecules per uc / 5-8 molecules per uc | Low loading: toluene in intersections. High loading: toluene in the intersection and the straight channels. | Huang et al.[31] |
| **Adsorption isotherm (TG-GSC)** | Toluene (pure) | Sample: H-MFI Si/Al = 22.5 Loadings: up to 4 molecules per uc | Up to 1 toluene molecule at each of the 4 intersections in a uc | Zheng et al.[27] |
| **Synchrotron PXRD** | Toluene/water (up to 45 mg/L) | Sample: H-MFI Si/Al = 140 Loading: 6.4 molecules per uc | 2 types: TOL1 in the straight channels, TOL2 in the intersection. | Rodeghero et al.[65] |
| **PXRD**[B] | Toluene/Hexane binary mixture in water | Sample: H-MFI Si/Al = 140 Loading: 0.8 toluene and 7.2 hexane molecules per uc | TOL1: in the straight channels. HEX1: 4 molecules in the intersection HEX2: 3.2 molecules in the sinusoidal channels | Rodeghero et al.[72] |
| **SCXRD**[A] | p-nitroaniline (pure) | Sample: H-MFI Loading: 4 molecules per uc | The molecules are located in the intersection and loosely connected through the straight channels. | Koningsveld et al.[73] |
| **SCXRD**[A] | p-dichlorobenzene (pure) | Sample: H-MFI Loading: 2.56 molecules per uc | Adsorbate localized in the intersections | Koningsveld et al.[71] |
| **SCXRD**[A] | Naphthalene (pure) | Sample: H-MFI Loading: 3.68 molecules per uc | Adsorbate localized in the intersections | Koningsveld et al.[70] |
| **SCXRD**[A] | Benzene (pure) | Sample: Silicalite-1 Loading: 7.2 molecules per uc | Intersection and the straight channels. | Kamiya et al.[74] |
| **SCXRD**[A] | Phenol (pure) | Samples: Silicalite-1 and H-MFI Si/Al = 140 Loading: 8 molecules per uc | Intersection and the sinusoidal channels | Kamiya et al.[34] |
| **SCXRD**[A] | Linear molecules: 1-butene<br>Bent molecules: n-butane, n-pentane, n-hexane<br>Branched molecule: isopentane | Sample: Silicalite-1 | Low loading: linear molecules in the straight channels. Bent molecules in the sinusoidal channels and branched in the intersection<br>High loading: linear and bent molecules in both channels. Branched molecules in the intersection and sinusoidal channels. | Fujiyama et al.[33] |
| **SCXRD**[A] | p-xylene (pure) | Sample: Silicalite<br>Loading: 8 molecules per uc | p-xylene located in the intersection and sinusoidal channels | Hwang[75] |
| **Sorption isotherm (gravimetry)** | Ethane, propane, n-butane (pure) | Sample: H-MFI; Si/Al = 138 | Sorption to sinusoidal channels for loadings < 8 molecules per uc Molecules in excess to that are located in the intersections | Richards and Rees[76] |
| **Grand Canonical Monte Carlo simulations** | Benzene, p-xylene (pure) | Sample: Silicalite | Sorption to intersections for loadings < 4 molecules per uc Molecules in excess to that are located in the sinusoidal channels | Snurr et al.[21] |
| **Molecular Mechanics** (Dreiding FF) | Benzene, Toluene, p-xylene | Sample: Silicalite | Lowest energy position in the intersection | Nakazaki et al.[77] |

[A] Single crystal X-ray diffraction. [B] Powder X-ray diffraction.

Despite the commercial relevance of aqueous-phase catalysis and solvent-modulated reactions, a molecular-level understanding of competitive adsorption between water and aromatics in zeolites remains incomplete with critical knowledge gaps in the role of framework composition, adsorption sequence, and local heterogeneities in governing site-specific interactions and final distribution of adsorbates. Investigating competitive adsorption, toluene and water represent two chemically and physically contrasting species. Toluene, a weakly polar aromatic hydrocarbon, interacts predominantly via dispersion and π interactions, while water exhibits strong affinity for BAS and silanol groups through hydrogen bonding and dipolar interactions. Water can outcompete organic molecules for access to polar and acidic sites, which in highly siliceous MFI are preferentially located in channel intersections.[27] The distinctly different adsorption behavior of toluene and water, their immiscibility in bulk and the enhanced solubility of toluene in nanoconfined water[78] render the toluene–water pair an interesting probe to study competitive adsorption in conditions relevant to the industrially relevant aqueous-phase catalysis, biomass upgrading, and solvent-modulated reactions. The present study illustrates how solid-state NMR can be used to give further insights into competitive adsorption phenomena in zeolites, supporting the localization of guest molecules. The limitation of this approach is also discussed, and we point out that multi-diagnostic investigation of complex systems is critical to understand the host-guest interactions in porous systems. As a concrete example, the competitive adsorption of toluene and water in MFI zeolites with varying Si/Al ratios is investigated using advanced 1D and 2D solid-state NMR. Systems with varying adsorbate loadings were investigated to decipher the influence of framework polarity and pore topology on the distribution of water and toluene throughout the MFI pores under loading of toluene in presence or absence of water.

## 2. Experimental Methods

### 2.1 Materials

Three commercial $NH_4$-form MFI-type zeolites exhibiting Si/Al ratios of 11.5 (CBV2314), 25 (CBV5524), and 140 (CBV28014) were obtained from Zeolyst and converted to their acid form by calcination at 550 °C for 6 h with a temperature ramp of 1 °C.min$^{-1}$. The final H-form samples are dubbed MFI-11.5, MFI-25, and MFI-140. Toluene was procured from Sigma-Adrich (99.8% reagent-grade). Hexane was acquired from VWR (99% purity), and furfuryl alcohol from Sigma-Aldrich (98% purity). All water used was ultra-pure (Sartorius™ Lab water type 1). All chemicals were used without further purification.

### 2.2 Characterization

<u>ICP:</u> Inductively coupled plasma optical emission spectrometry (ICP-OES) was performed on an axial simultaneous Varian 720-ES with cooled cone interface and oxygen free optics to characterize the Si and Al content of the MFI samples and compare it to the nominal composition provided by Zeolyst. <u>Physisorption:</u> $N_2$ adsorption-desorption isotherms were measured with at -196 °C using a Micromeritics® TriStar II 3020 after the samples were degassed at 150 °C. From the isotherms, the BJH (Barrett–Joyner–Halenda) micropore volume and the BET (Brunauer–Emmett–Teller) specific surface area of the zeolites was derived.

NMR spectroscopy: NMR measurements were performed on a Bruker Avance III 500 MHz (11.7 T) spectrometer equipped with a 4 mm solid-state H/X/Y triple channel MAS probe. $^1$H direct excitation spectra were acquired at spinning rates up to 15 kHz.[37] The $^1$H excitation employed π/2 radio-frequency (RF) pulses at 96 kHz, averaging 8 transients with a recycle delay of 5 s. $^1$H-$^1$H EXSY experiments employed a pulse sequence with mixing time of 20 ms to probe the exchange dynamics between the BAS sites and adsorbed toluene. $^1$H and $^{13}$C spectra were referenced using adamantane $^1$H and $^{13}$C (methylene) resonances at 1.81 ppm and 38.5 ppm, respectively, and $Q_8M_8$ for $^{29}$Si at -108.36 ppm, both as secondary references relative to tetramethylsilane (TMS) at 0 ppm. Spectral decomposition was performed using mixed Gaussian-Lorentzian line shapes to resolve overlapping components using DMFIT software.[79] Spectral decomposition of the $^1$H NMR spectrum of the dry zeolite samples enabled quantification of BAS. Framework Si/Al ratios were determined via $^{29}$Si NMR by spectral decomposition and integration of $Q_{4(0Al)}$ and $Q_{4(1Al)}$ species. The NMR derived data agrees with ICP-OES results and the nominal values provided by Zeolyst. $^1$H–$^1$H radio-frequency driven recoupling (RFDR) correlation experiment was done at 10 kHz MAS frequency using 90° pulses of strength 96 kHz with a mixing period of 4.8 ms. 16 scans were collected in each slice with an increment of 66.6 µs for a total of 1024 slices for the 2D experiment.

A summary of the physicochemical characteristics of the MFI zeolites investigated here is shown in **Table 2**.

**Table 2**. Chemical composition, micropore volume ($V_{micro}$), BET specific surface area ($A_{BET}$) and specific external surface area ($A_{ext}$) of the zeolite samples as characterized by ICP-OES, NMR spectroscopy and $N_2$ adsorption-desorption isotherms.

| Sample | Chemical composition | | | | Surface and porosity analysis | | |
|---|---|---|---|---|---|---|---|
| | Si/Al[a] | Si/Al[b] | BAS[c], A | Framework Al[d] | $V_{micro}$, B | $A_{BET}$, C | $A_{ext}$, C |
| **MFI-11.5** | 11.7 | 14.4 | 1.058 | 0.902 | 0.157 | 365 | 22.0 |
| **MFI-25** | 27.7 | 28.3 | 0.672 | 0.545 | 0.166 | 402 | 36.9 |
| **MFI-140** | 135 | 133 | 0.202 | 0.121 | 0.186 | 368 | 12.2 |

[a] from ICP-OES. [b] from $^{29}$Si NMR. [c] from $^1$H qNMR. [d] from $^{27}$Al NMR. **A**: mmol.g$^{-1}$. **B**: cm$^3$.g$^{-1}$. **C**: m$^2$.g$^{-1}$.

### 2.3 Solid-state NMR guided adsorption experiments

For the adsorption experiments and further NMR characterization, the zeolite samples were packed in 4 mm $ZrO_2$ NMR rotors and dried for 16 h at 200 °C at 75 mTorr. The liquid adsorbates (toluene and/or water) were dosed to the samples in the rotors with a Hamilton® chromatography syringe of 10 µL. Following this addition, the rotor was closed and equilibrated overnight at 60 °C to load the sample with the adsorbate. This procedure has previously been shown to lead to a homogeneous distribution of water and linear alcohols ($C_1$-$C_5$) throughout the pore system of MFI-type zeolites.[32,37] After equilibration, the samples were subjected to NMR characterization. The dosage of the adsorbate was increased stepwise up to a maximum equivalent to 120% of the BET micropore volume ($V_{BET}$) of the samples.

### 2.4 Selective pore blocking

Previous research using confocal fluorescence imaging spectroscopy and absolute quantification via NMR spectroscopy has shown that the oligomerization of furfuryl alcohol can be promoted to selectively block either the straight or sinusoidal channels of MFI zeolites.[53] Acid-catalyzed furfuryl alcohol oligomerization in presence of water led to selective occupation of furfuryl alcohol in the straight channels, while the presence of non-polar solvents such as dioxane or 2-butanone favored polymerization in the sinusoidal channels. This reaction was used here to selectively promote oligomerization of furfuryl alcohol in the straight pores of an MFI-140 zeolite, blocking the accessibility of toluene to the pore network, while leaving accessible only the segments of sinusoidal pores in the extremities of the zeolite crystals. Samples of MFI-140 with selectively filled straight channels were prepared by treating the zeolite with a furfuryl alcohol–water mixture (1:10 v/v), followed by centrifugation and washing with water.[53] An additional sample where complete pore filling with furfuryl oligomers was achieved by polymerization of a 1:4 v/v mixture. The spent catalyst was then washed with water, dried and packed into an 4 mm MAS rotor. The sample packed in rotor was further vacuum-dried and subjected to addition of toluene (Section 2.3).

## 3. Results and Discussion

### 3.1 Single component adsorption of toluene on MFI-11.5, MFI-25 and MFI-140

#### 3.1.1 $^1$H NMR detected adsorption profiles

To investigate the influence of framework polarity on toluene adsorption in MFI zeolites, $^1$H NMR detected toluene adsorption was performed on zeolites with varying Al-content (MFI-11.5, MFI-25 and MFI-140) as function of toluene loading (**Figure 2** and **Table S1**). The $^1$H NMR spectra enable to differentiate adsorbed toluene species based on their local environments, mobility, and interaction strength with the framework. Spectral features such as chemical shift, line width, and the presence or absence of *J*-coupling fine structure provide insights into the spatial organization and dynamic behavior of the adsorbate molecules. Broadening of the $^1$H resonances typically indicate strongly bound or confined molecules with restricted mobility. This results in enhanced dipolar relaxation and shortens the transverse relaxation times ($T_2$), thus leading to broad resonances. Weakly adsorbed, mobile toluene species in contrast display narrower lines. At higher loadings, when excess adsorbate molecules, i.e. toluene, exist in a liquid-like state, very sharp resonances with observable through-bond scalar *J*-coupling patterns may appear,[32] indicative of isotropic tumbling and minimal interaction with the zeolite surface.

The $^1$H NMR spectra of the dehydrated zeolite samples, all pores void of any adsorbate, revealed the presence of $^1$H in silanols at 1.8 ppm, aluminols at 2.55 ppm and BAS at 3.9 ppm (**ESI Fig. S1**). A broad resonance observed between 4 – 10 ppm is attributed to chemical distribution of BAS species induced by for example, H-bonding, the presence of silanol nests, etc. Upon adsorption of toluene, the $^1$H NMR spectra for the different MFI zeolites revealed considerably different NMR signatures as function of toluene loading, providing information on different adsorption regimes (**Figure 2**).

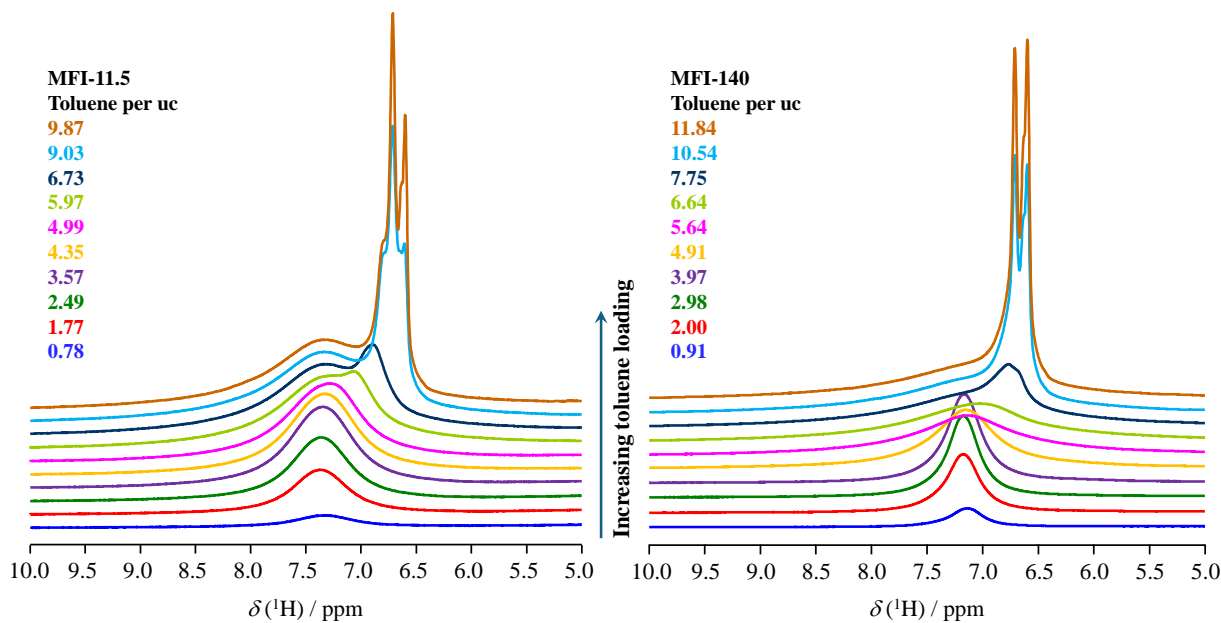

**Figure 2**. $^1$H MAS NMR spectra of dehydrated MFI-11.5 (left) and MFI-140 (right) following sequential adsorption of increasing amounts of toluene.

Overall, 3 regimes were observed for all samples, irrespective of their Si/Al ratio. In the case of MFI-11.5, at low toluene loadings, broad resonances appeared at 7.36 and 2.43 ppm. These resonances correspond to the aromatic and methyl protons, respectively (component C1). These signals exhibit significant downfield shifts compared to resonances of liquid toluene, appearing respectively at 6.9 – 7.1 ppm and 2.1 ppm.[80] Up to a loading of 4.4 molecules per unit cell ($V_{added}$ = 0.51 $V_{micro}$), the resonances increase in intensity in accordance with the increased toluene-loading. At loadings between 4.4 to 5 toluene molecules per uc ($V_{added}$ = 0.5 to 0.6 $V_{micro}$), slight broadening and a splitting in the toluene resonances was observed, with additional resonances appearing at ca. 7.2 and 2.28 ppm as a shoulder on top of the initial resonances of the aromatic and methyl protons, respectively. Above 6 molecules per uc ($V_{added}$ = 0.7 $V_{micro}$), an additional resonance (component C2) was observed with aromatic and methyl $^1$H chemical shifts at 7.08 and 2.1 ppm respectively. With further addition of toluene, this component exhibited a continuous upfield shift to respectively 6.8 and 1.84 ppm. While this resonance was noticeably narrower, indicating an increased mobility, it still lacked the fine structure expected for liquid-like toluene external to the zeolite pores. This indicates these molecules were still confined, but less than component C1. While the upfield shift of fraction C2, with respect to C1, could be attributed to chemical exchange effects, its final chemical shifts were higher than those for liquid toluene, indicating this toluene species is still in interaction with the zeolite surface. The enhanced mobility and chemical shift observed for this fraction suggest two possible adsorption locations for component C2: (i) the external surface, including pore mouths, and/or (ii) the channel intersections. Further experiments to distinguish between these two possibilities are described in **section 3.1.2**. Upon addition of more than 8 toluene molecules per uc ($V_{added} > 0.9\ V_{micro}$), extremely narrow resonances with *J*-coupling based splitting were observed (component C3) between 6.7 – 6.6 ppm. Similar resonances are observed for toluene nanodroplets dispersed in water (triplet – triplet – doublet pattern between 6.6 – 6.47 ppm) (**Figure 3**),[81] implying the chemical shift difference with respect to neat toluene (triplet – triplet – doublet pattern between 7.13 to 7 ppm) and toluene dissolved in $D_2O$ (triplet – doublet – triplet pattern between 7.21 – 7.1 ppm) results from electric field effects at the toluene water interphase.[82] Interestingly the *J*-coupling derived splitting patterns observed for toluene in the nanodroplets dispersed in $D_2O$ and toluene present in population C3 are similar and distinctly different from those observed for neat toluene and toluene dissolved in $D_2O$. Combined, these observations allow to assign the population C3 as toluene molecules adsorbed on the outer surface of the zeolite crystals. To further confirm the assignment of component C3, toluene was added to a MAS rotor containing α-alumina, a material devoid of any microporous framework. The resulting $^1H$ resonances (**Figure 3**) were predominantly sharp and appeared at chemical shifts corresponding to those of component C3.

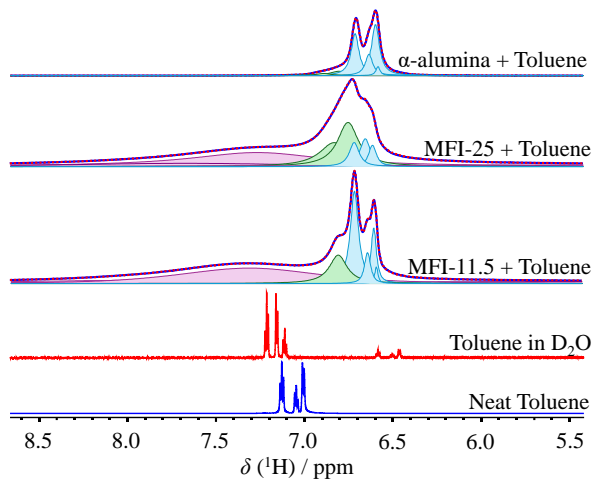

**Figure 3**. 1H NMR spectra of neat toluene, toluene in D2O (0.9 g/L), and toluene loaded in MFI-11.5 (9.9 per uc), MFI-25 (10.6 per uc), and in α-Alumina (29.8 mg toluene per gram) (purple represents C1, green represents C2, and blue represents C3).

Note that also some of the zeolite associated $^1H$ resonances change upon toluene addition (**ESI Fig. S2**). Upon progressive addition of toluene, an isosbestic point appeared as the BAS proton resonance broadened and shifted from 3.9 ppm to ca. 4.8 ppm (**ESI Fig. S2**). This is an indication of interaction with toluene, which was also confirmed by the appearance of a cross peak in the EXSY spectrum of MFI-11.5 loaded with 1.77 toluene molecules per uc (section 3.1.2, **ESI Fig. S3**), indicating interaction between the aromatic protons and the shifted and broadened BAS resonance. The silanol resonances at 1.8 ppm on the other hand did not exhibit any significant changes upon toluene addition.

In the case of MFI-140 (**Figure 2**), at loadings of 0.91 toluene molecules per uc, the resonance associated with component C1 was revealed to be an envelope signal consisting of two subpopulations: a minor contribution at 7.24 ppm overlapping with predominant resonance at 7.12 ppm (**ESI Fig. S4**). Based on the relative intensity between these two resonances, the populations can be estimated to be 0.26 and 0.65 toluene per uc respectively. The resonance at 7.24 ppm likely corresponds to aromatic $^1H$'s interacting with BAS, also confirmed by the transformation of the BAS resonance at 3.95 ppm into the broad resonance analogous to what happens in MFI-11.5. The decrease in intensity of the BAS at 3.95 ppm indicates that circa 24% BAS (0.28 BAS per uc) interacted with toluene, a number that is similar to the subpopulation of C1 at 7.24 ppm. At this toluene loading, the toluene/BAS ratio was 0.8. The fact that at this loading only a minor fraction of the adsorbed toluene is interacting with BAS indicates that interaction with the Brønsted acid site is not the predominant interaction driving toluene adsorption in high-silica zeolites. The major resonance exhibited a narrower line shape and chemical shift of 7.15 and 2.32 ppm for aromatic and methyl protons, respectively, upfield shifted compared to the C1 components in MFI-11.5. This indicates an interaction with lesser extent of electronic changes leading to lower chemical shift changes compared to bulk-like toluene, predominantly dispersion-type interaction. The resonances were also considerably narrower (190 Hz) compared to those observed for toluene adsorbed in MFI-11.5 (335 Hz), thus indicating either higher mobility or reduced chemical shift broadening effects induced by slight differences in local order at adsorption sites. The resonance associated with the BAS site resonance completely disappeared at ca. 2 toluene molecules per uc ($V_{added} = 0.20\ V_{micro}$), corresponding to a toluene/BAS ratio of ca. 1.7. The silanol resonance, however, retained its original intensity, indicating that interaction the silanol protons do not take part in the adsorption.

In the case of MFI-25, component C3, observed at loadings above 10 toluene molecules per uc ($V_{added} > 1.0\ V_{micro}$), exhibited a significantly broadened line shape devoid of scalar coupling fine structure, suggesting significantly stronger interactions of toluene with the external surface of MFI-25. Note that $N_2$ physisorption measurements (**Table 1**) indicated a substantially higher external surface area for MFI-25 compared to the other MFI sam-ples. This indicates the particle size of the MFI-25 crystals was much smaller, a feature which could form the basis for the observed broadening.

### 3.1.2 Assignment experiments

In **section 3.1.1,** toluene component C1 was attributed to toluene confined in the zeolite pores, while two possible locations for component C2 were suggested: (i) adsorption to the external surface (including to the pore mouth), and/or (ii) adsorption in the intersections between the straight and sinusoidal channels. To distinguish these two options, two additional experiments were performed. For MFI-11.5 loaded with 6.73 toluene per uc, for which component C3 is visible, a $^1$H RFDR NMR spectrum was acquired (**Figure 4**). In a $^1$H RFDR spectrum, diagonal peaks correspond to individual $^1$H resonances, whereas cross-peaks (off-diagonal contours) arise from through-space magnetization transfer mediated by dipolar couplings between distinct proton sites. In the spectrum of **Figure 4**, a prominent cross-peak between component C1 and C2, highlighted in yellow, can be seen. The presence of this cross-peak confirms that C1 and C2 are in close spatial proximity within the structure, implying that any adsorption of C2 on the external surface must occur at the pore mouths rather than on remote surface sites. This restricts component C2 to two localization scenarios: adsorbed in the pore mouth or in the channel intersections.

To further distinguish these options, furfuryl alcohol oligomerization in presence of water was exploited to selectively occupy the straight channels and intersections in MFI-140 (**ESI Fig. S5**), implementing conditions previously reported by our group.[53] This blocks access to the pore system for toluene, leaving predominantly the pore mouths and external surface accessible for toluene adsorption. The $^1$H NMR spectra of this sample before and after toluene addition are shown in **Figure 5**. Quantitative $^1$H NMR spectroscopy of the dehydrated, spent catalyst (blue trace in **Figure 5**), allowed to quantify the loading of furfuryl alcohol oligomer in the pore system at 0.14 mg/g zeolite. Using the bulk density of furfuryl alcohol, the volume occupied by the oligomer in the MFI pores after washing with water was estimated to be 73 % of micropore volume, in line with the expected polymerization of furfuryl alcohol in straight channels and intersections.[53] Addition of toluene to this sample at a dosage of 1 toluene molecule per uc revealed components C1 and C2 in 0.6 and 0.4 per uc respectively. The appearance of C2, even before full filling of micropore volume may be attributed to diffusion limitations imparted by the tortuous zigzag channels. Increasing the toluene loading to 3 toluene molecules per uc, liquid-like resonances (C3) appeared, superimposed on the C2 signal. This implies that the preferred adsorption sites for toluene (C1) are already occupied by the furfuryl alcohol oligomers or difficult to access due to the presence of the polymer, which cannot be displaced by toluene. To further confirm the assignment, MFI-140 zeolite was prepared under conditions to achieve full pore filling by oligomers of furfuryl alcohol by a modified procedure (see section **Experimental Methods).** The spent catalyst after rinsing with water contained 37 mg furfuryl alcohol oligomers per g zeolite, equivalent to 150 % of its pore volume (**ESI Fig. S5**). Addition of toluene (1 per uc) led to appearance of a minor fraction of C1, but immediately reaching C2 and C3. Small volumes of channel segments at the extremities of the crystals, still accessible after furfuryl alcohol polymerization, account for the appearance of traces of component C1. The appearance of component C2 even when channels and intersections are blocked confirms C2 as pore mouth adsorbed toluene molecules.

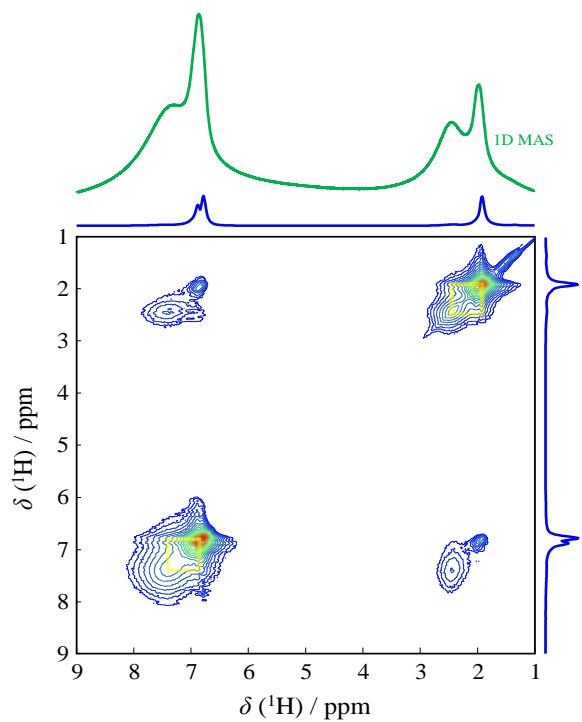

**Figure 4**. $^1$H-$^1$H RFDR MAS NMR spectrum of MFI-11.5 loaded with 6.73 toluene molecules per uc The yellow square highlights the correlation between toluene components C1 and C2.

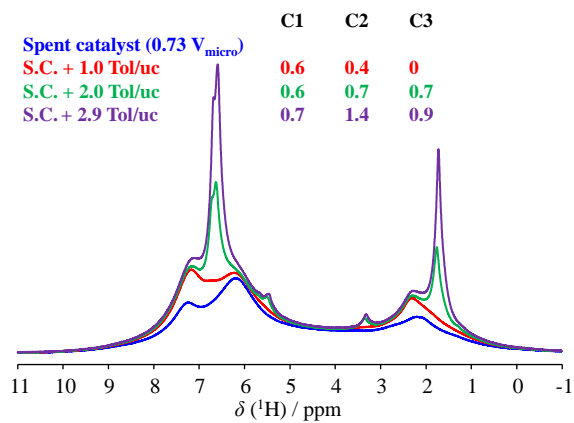

**Figure 5**. $^1$H MAS NMR spectra of pore-inaccessible MFI-140 before toluene loading (blue) and after the addition of 0.95 (red), 1.96 (green), and 2.89 (purple) of toluene molecule per uc to this sample with blocked pores.

### 3.2 Competitive adsorption of water and toluene

Having localized toluene fractions adsorbed onto the zeolite crystals, the competitive adsorption of toluene and water was investigated by monitoring the evolution of toluene components C1, C2, and C3 upon stepwise water addition. The competitive adsorption was examined by first filling a dehydrated MFI-140 with 8.2 toluene molecules per uc, a loading at which the MFI channels are saturated and a fraction of the toluene is already adsorbed to the pore mouths, while bulk-like toluene is still absent (see **Table S1** and **Figure 6**). Subsequently, water was added incrementally to the toluene-loaded sample. After the first water addition ($V_{added} = 0.07\ V_{micro}$), water already partially displaced toluene component C1. The desorbed toluene re-adsorbed primarily onto available pore mouth sites, as seen in **Figure 6** and the inset on its left side. The fact that water preferentially displaced confined toluene rather than adsorbing to easily accessible pore mouths indicates that water adsorption happens on sites with strong hydrophilicity. Increasing the total added water volume to $0.32\ V_{micro}$ resulted in the desorption of 2.5 toluene molecules per uc from component C1 while also component C2 was partially desorbed, with the desorbed species transforming into liquid-like toluene (component C3). To rationalize the desorption of C1, we note that MFI-140 contains BAS, silanol, and aluminol groups at concentrations of 1.2, 0.5, and 0.07 per uc, respectively, totaling 1.74 hydrophilic sites per uc While this number is considerably lower than the amount of displaced C1 toluene molecules, water has a strong preference for interacting with these hydrophilic sites, especially BAS. Such interactions perturb the local environment in the channels and indirectly destabilize confined toluene, promoting its desorption. Although toluene does not typically interact strongly with silanol groups, their hydration will still contribute to the overall disruption of confined adsorption. These observations suggest that water–framework interactions at hydrophilic sites, rather than channel topology, play the dominant role in driving the desorption of confined toluene.

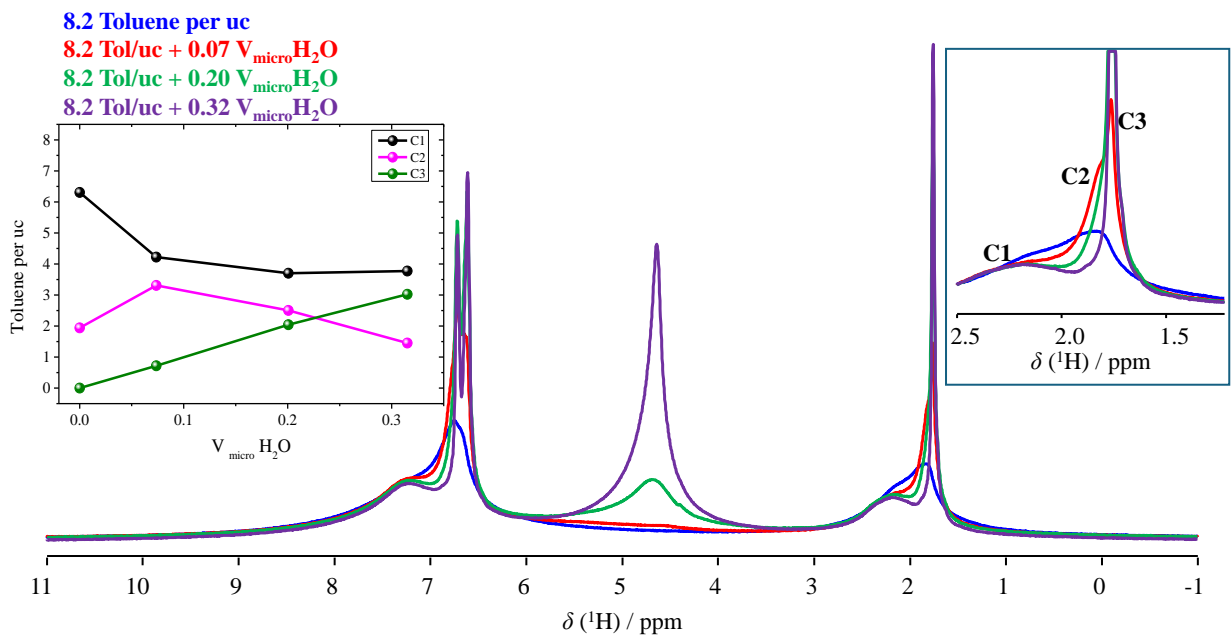

Figure 6. $^1$H MAS NMR spectra of MFI-140 initially loaded with 8.2 toluene molecules per uc (blue) and subsequently subjected to increasing amounts of water addition. Inset (right) shows a magnified view of the methyl region, highlighting the redistribution of toluene among C1, C2, and C3 populations. Inset (left): Quantitative analysis of the evolution of each component as a function of water loading.

### 3.3 Need for *in situ* multi-diagnostic methodologies

Whereas NMR spectroscopy can differentiate between toluene confined in the zeolite channels, toluene adsorbed to pore mouths and toluene at the outer surface of MFI zeolites, resolving the precise location of toluene molecules confined within the MFI pore system remains challenging. These confined subcomponents manifest in $^1$H NMR as a degenerate spectral feature (C1). A possible route to distinguish between these different subcomponents would be to integrate the NMR data with orthogonal information acquired from complementary techniques such as X-ray diffraction, Raman spectroscopy, adsorption measurements and modelling. However, the use of multiple techniques introduces its own difficulties to match and consistently interpret data acquired under the often-different experimental conditions imposed by the sample environment of the complimentary methodology. For many samples, even small variations in conditions (pressure, temperature, water content, etc.) can lead to markedly different behavior. This probably contributes to the myriads of often inconsistent toluene site assignments reported in literature (infra). This challenge underscores the need for advanced multi-diagnostic methodologies, ideally performed *in situ* within the same equipment, or across different sample environments linked by a common metric to assist later integration. Recently, we reported the development of in situ multi-diagnostic NMR/impedance spectroscopy, enabling simultaneous measurement of NMR spectra and sample impedance.[83] Similarly impedance spectroscopy and conductivity measurements were integrated into a sample environment for high-pressure X-ray scattering and spectroscopy.[84] While not yet experimentally realized, also the integration of Raman spectroscopy is technically feasible. It offers the possibility of virtually integrating NMR spectrometers and large-scale X-ray and neutron facilities using both impedance and Raman spectroscopy as a common diagnostic, enabling reliable integration of results into a consistent multidimensional dataset. Realization of such *in situ* or virtually integrated multi-diagnostic methodologies would enable rigorous investigation of complex host–guest interactions.

## 4. Conclusions

Quantitative $^1$H MAS NMR spectroscopy reveals toluene adsorption in dehydrated MFI zeolites to evolve with loading, reflecting progressive adsorption of toluene at different locations in the framework. The observed behavior is independent of the Si/Al ratio in the range between 11.5 and 140. Overall, 3 spectral components, and thus 3 different fractions of toluene were observed. While interaction of toluene with the Brønsted acid sites is observed, this interaction is not driving the adsorption. A series of 1D and 2D NMR experiments, combined with sample engineering with pore blocking with furfuryl oligomers enabled assignment of these components to, respectively, toluene confined in the MFI channels, toluene adsorbed in the pore mouths and toluene at the outer surface of the crystals. Confined toluene (C1) exhibits significantly broadened $^1$H NMR resonances and it is the only component observed at loadings lower than 5 toluene molecules per uc. Increasing the toluene loading from 0 to 5 toluene molecules per uc, the pore occupancy and thus also nuclear dipolar interactions between toluene molecules intensify. This is reflected in broadening of the NMR resonances for population C1 and reveals close proximity between toluene molecules in the zeolite channels. Above 6 molecules per uc, toluene adsorbed to pore mouths appear as component C2, with characteristics of increased mobility, but exhibiting broadening with respect to the third population (C3), which appears at loadings beyond 8 toluene molecules per uc and is assigned as toluene adsorbed on the outer surface of the crystals. Both the C2 and C3 populations exhibit broadening with respect to bulk, liquid toluene. Upon water addition to MFI-140 previously loaded with toluene, competitive adsorption between water and toluene takes place, with water adsorbing selectively to the Brønsted acid sites and the silanol nests in defects, partially displacing confined toluene to the pore mouths. Increasing water loading causes displacement of toluene from the pore mouths to the outer surface and would directly impact acid catalysis processes. Resolving the location of confined molecules within the MFI pore network remains challenging and assignments should always be accompanied by detailed sample characterization to guide data interpretation and to allow reconcile results with the other reports in literature. Complete understanding of complex host-guest interactions in porous materials demands multi-diagnostic characterization of both the ordered and disordered molecular arrangements and also of physico-chemical phenomena occurring across multiple time and length scales. This can only be achieved when reliable metrics exist across multiple methods which can serve as a reference to benchmark different samples and sample conditions.

**Acknowledgements**

NMRCoRe acknowledges the Hercules Foundation for infrastructure investment (AKUL/13/21), and the Flemish government (Department EWI) for financial support (I001321N: Nuclear Magnetic Resonance Spectroscopy Platform for Molecular Water Research) and infrastructure investment via the Hermes Fund (AH.2016.134). This work has received funding from the European Research Council (ERC) under grant agreement no. 834134 (WATUSO) and was supported by the Centre for Molecular Water Science (CMWS) in an Early Science Project. The authors thank Gina Vanbutsele (KU Leuven), Wauter Wangermez (KU Leuven), and Loes Verheyden (KU Leuven) for their assistance with sample characterization.